# Verification of crystal collimation model in experiment

V.M. Biryukov[♦]

*Institute for High Energy Physics, Protvino, 142281, Russia*

**Abstract**

The studies of crystal collimation in the experiments at Relativistic Heavy Ion Collider and Tevatron and in computer simulations reveal strong coherent effects observed in a very broad angular range. Our theory explains the effects by coherent scattering on the potential of bent crystal atomic planes, which amplifies beam diffusion in accelerator by orders of magnitude. This coherent scattering in bent crystal is being studied in a CERN SPS experiment. We present Monte Carlo predictions for the SPS and Tevatron experiments, and show the implications of the coherent scattering effect for crystal collimation in the Large Hadron Collider.

**1. Introduction**

The studies of beam extraction and collimation assisted by channeling in a bent crystal have well progressed in recent decades [1-12]. The technique was applied at up to world highest energy [2,6], in good agreement with predictions [13]. In IHEP Protvino, much of the physics program relies on crystal channeled beams used regularly since 1989 [12]. Crystal systems extract 70 GeV protons from IHEP main ring with efficiency of 85% at intensity up to $4\times10^{12}$ protons [12].

The use of crystal for collimation purpose was first proposed for 3 TeV UNK [14] and 20 TeV SSC [15] colliders in 1991. A bent crystal, serving as a primary element, should coherently bend halo particles onto a secondary collimator. A proof-of-principle demonstration experiment of this kind was performed in 1999 at IHEP where a factor-of-2 reduction in the 70-GeV proton accelerator background was obtained with a bent crystal incorporated into beam cleaning system [16].

The first experiment on crystal collimation with heavy ions was done at the Relativistic Heavy Ion Collider [9-11]. A 5-mm *Si* crystal collimator deflected *Au* ions of 250 GeV per unit charge onto a copper scraper. RHIC crystal collimator efficiency measured for gold ions as a function of the crystal angle was found in good agreement with simulations with the measured

---

[♦] http://mail.ihep.ru/~biryukov/

machine optics. For the 2003 RHIC run, the average measured efficiency was 26%, dominated by a high angular spread of the beam that hits the crystal face as set by machine optics. Crystal extraction of *Pb* ions was earlier demonstrated at CERN SPS with efficiency of 4-11% for a long (40 mm) Si crystal [3].

It would be promising to apply the bent-crystal technique for a beam halo scraping at high-energy colliders such as the Tevatron [17] and LHC [18] where an order of magnitude reduction is expected in the accelerator-related backgrounds. Roughly, the idea of crystal collimation is that crystal traps 90-95% particles and channels them well into the depth of a collimator. The remaining 5-10% particles are not channeled and are handled in traditional way by a multi-stage "amorphous collimation" system.

The findings at RHIC [9-11] and recently started crystal collimation experiment at the Tevatron [19] strongly suggest that the picture is more complex than that. The Monte Carlo study [20] of the particle dynamics in single and multiple interactions with a bent crystal in different angular ranges in the environment of collimation experiment has shown that coherent scattering of nonchanneled particles in bent crystal causes a strong diffusion in the plane of betatron oscillations in the accelerator ring, affecting the collimation.

## 2. Predictions for CERN SPS experiment on volume reflection

The difference in behavior of nonchanneled particles in a bent crystal and in amorphous body is the subject of a new experiment at CERN SPS [21] which is a step towards the application of crystal collimation at the LHC, making use of H8 external micro-beam line with unique possibilities for crystal tests. Here we provide Monte Carlo predictions for this experiment in order to verify the computer model [22,23] earlier used for predictions of many of experiments [1-12] and for the LHC [18,24]. We took into account the characteristics of the bent Si(111) crystal chosen for the test (quasimosaic type [25], 1 mm along the beam, bent 0.1 mrad), beam (3 μrad r.m.s.) and experimental setup [21] with scattering in the counters and beam windows. Like in the experiment, we select only protons incident within 0.1 mm from the crystal center in the horizontal plane. Figs. 1-2 show 400 GeV proton beam angular distributions downstream of the crystal. Fig. 1 shows channeling for best alignment of the crystal where we expect 69% protons bent the full angle of 0.1 mrad.

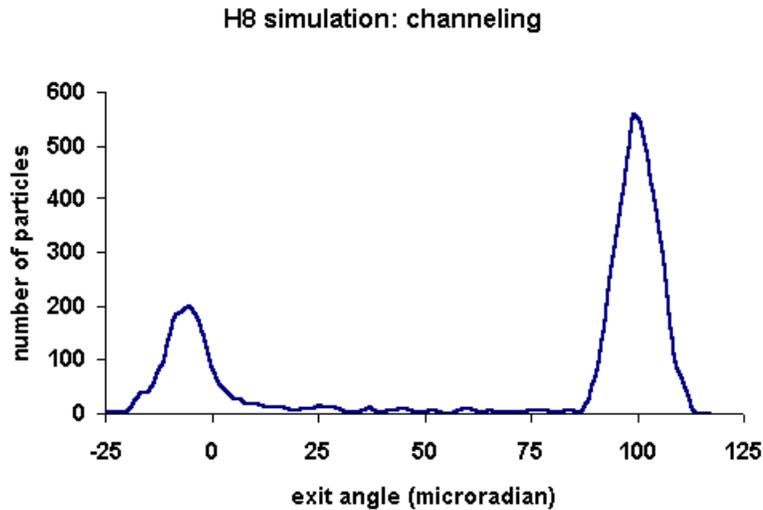

**Figure 1** Angular distribution of 400 GeV protons downstream of the aligned crystal.

Very interesting is the case where crystal is tilted a fraction of its bend angle, i.e., between 0 and 0.1 mrad, to the incident beam. Fig. 2 shows that 95% protons exit from the crystal with horizontal angle x'<0, and only 5% with x'>0. The exit beam consists of a major peak containing 97.4% protons that is shifted in angle ("reflected") by 13 µrad (most probable angle) w.r.t. the incident beam, and a long flat tail (2.6% protons) of volume-captured protons. This angular shift is known as "volume reflection" [26] caused by a coherent scattering on the potential of bent atomic planes. In our case, the width of the reflected peak is about the same as the beam width at a random alignment, so one can say that a beam is reflected practically undistorted, with efficiency of reflection of ~97.4%.

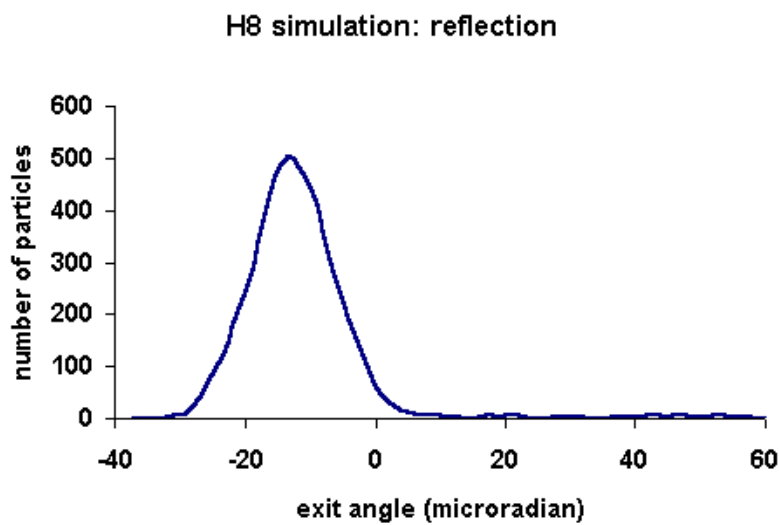

**Figure 2** Angular distribution of 400 GeV protons downstream of the crystal angled at 50 µrad to the incident beam.

## 3. Predictions for the LHC and Tevatron crystal collimation

The contribution from coherent effects to overall scattering in crystal is essential for collimation [20] therefore its measurement is very important. In order to find how this effect extrapolates to the LHC, we simulated the 7 TeV proton interaction with the crystals considered earlier [18] for collimation in the LHC. Fig. 3 shows an angular distribution downstream of a 5 mm Si(110) crystal bent 0.1 mrad. Under reflection conditions, the most probable exit angle is about -3 μrad, and the rms angle is 2.5 μrad, factor of 5.4 greater than in amorphous Silicon. The effective radiation length for scattering is just $L_R \approx 3.2$ mm, i.e., 5-mm Si crystal scatters like a 5-mm $W$ target! Crystal behaves like a "smart material" with very short $L_R$. This causes a strong diffusion in the plane of betatron oscillations in the LHC ring.

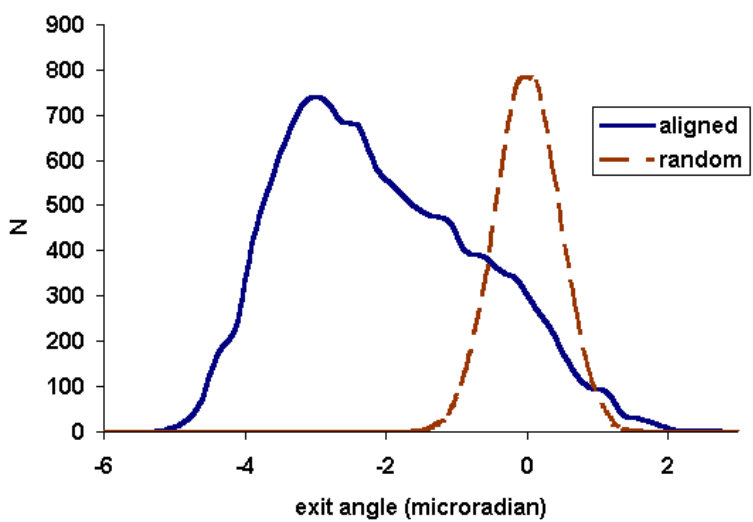

**Figure 3** Exit angular distributions for 7 TeV proton aligned to reflection and at random alignment.

We evaluate this diffusion in the LHC in the same way as been done at RHIC [9-11] and is under study at the Tevatron [19] experiments. The rate of nuclear interactions in the crystal is obtained as a function of a crystal alignment over a broad angular range, larger than the crystal bending angle. The LHC settings in this simulation were as in ref. [18]. The crystal was placed at 6σ and served as a primary element in collimation scheme. The secondary collimator was placed at 7σ. Particle tracking in the LHC lattice is done with linear transfer matrices. Each particle was allowed to make an unlimited number of turns in the ring and of encounters with the crystal until a particle either undergoes a nuclear interaction in the crystal or hits the secondary collimator (either because of a bending effect in the channeling crystal or because of the scattering events). A non-channeling amorphous layer 1 micron thick was assumed on the crystal surface due to its irregularity at a micron level.

As an example, we took a 5 mm Si(110) crystal bent 0.1 mrad, earlier considered in detail in the studies of LHC crystal collimation [18]. The simulated nuclear interaction rate in the crystal is shown in Fig. 4. The remarkable dip on the plot, about 98% down from the rate observed at random orientation, is due to channeling with very high efficiency predicted for the LHC. The dip is ~10 µrad wide, which is greater than 2 critical angles because here channeling is essentially multipass, multiturn effect. Protons encounter crystal several times, scatter, go on circulating in the ring, and then get channeled on some later encounter. This scattering contributes sizably to the width of the peak. Another distinct feature is the plateau on the plot, a 12% reduction in the rate because of the increased beam diffusion due to coherent scattering. The plateau width corresponds to the crystal bending angle, 0.1 mrad. Qualitatively, this plot is very similar to the plots measured at RHIC [9-11] and Tevatron [19] with a 0.44-mrad bent crystal.

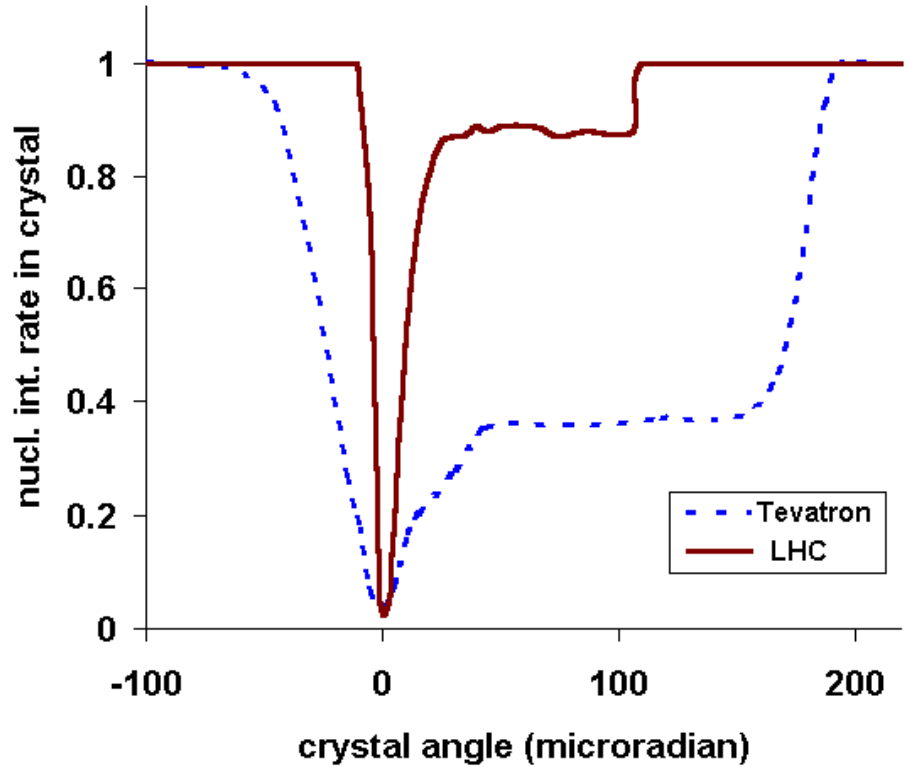

**Figure 4** The predicted crystal nuclear interaction rate for 980 GeV and 7 TeV protons in the collimation setups of the Tevatron and LHC.

Apart from checking theory in a single interaction of particle with a bent crystal, it is essential to check theory in the environment of a collimation experiment, with multiturn dynamics in the accelerator ring and multiple interactions with the crystal. Such a test is possible in the Tevatron, where a new bent crystal (Si (110), 3 mm along the beam, and 0.15 mrad bent) has been recently installed and tests should be started soon [19]. Our predictions for the Tevatron are presented in

Fig. 4. The rate suppression of 65% at plateau, 0.15 mrad wide, is expected for the new crystal in the Tevatron setup, with crystal at 5σ and the secondary collimator at 5.5σ. The expected channeling dip at crystal best alignment is about 96%.

## 4. Summary

The studies of crystal collimation renewed the interest to basic mechanisms of beam interaction with a bent crystal predicted decades ago. Our model predicts that the factor behind the new coherent effects in crystal collimation is a strong increase in the *mean square* angle of a particle scattered off the coherent field of a bent crystal. The tests of the model in the experiments in external beamline at CERN SPS and in accelerator ring in the Tevatron are essential for secure design of future applications of crystals and other channeling materials [27] at accelerators.

**Acknowledgements**

Support from INTAS-CERN grants 03-51-6155 and 05-103-7525 is gratefully acknowledged.